\PassOptionsToPackage{unicode}{hyperref}
\PassOptionsToPackage{hyphens}{url}
\documentclass[
]{article}
\usepackage{amsmath,amssymb}
\usepackage{iftex}
\ifPDFTeX
  \usepackage[T1]{fontenc}
  \usepackage[utf8]{inputenc}
  \usepackage{textcomp} 
\else 
  \usepackage{unicode-math} 
  \defaultfontfeatures{Scale=MatchLowercase}
  \defaultfontfeatures[\rmfamily]{Ligatures=TeX,Scale=1}
\fi
\usepackage{lmodern}
\ifPDFTeX\else
\fi
\IfFileExists{upquote.sty}{\usepackage{upquote}}{}
\IfFileExists{microtype.sty}{
  \usepackage[]{microtype}
  \UseMicrotypeSet[protrusion]{basicmath} 
}{}
\makeatletter
\@ifundefined{KOMAClassName}{
  \IfFileExists{parskip.sty}{%
    \usepackage{parskip}
  }{
    \setlength{\parindent}{0pt}
    \setlength{\parskip}{6pt plus 2pt minus 1pt}}
}{
  \KOMAoptions{parskip=half}}
\makeatother
\usepackage{xcolor}
\usepackage[margin=1in]{geometry}
\usepackage{color}
\usepackage{fancyvrb}

\DefineVerbatimEnvironment{Highlighting}{Verbatim}{commandchars=\\\{\}}
\usepackage{framed}
\definecolor{shadecolor}{RGB}{248,248,248}
\newenvironment{Shaded}{\begin{snugshade}}{\end{snugshade}}

\newcommand{\AttributeTok}[1]{\textcolor[rgb]{0.13,0.29,0.53}{#1}}

\newcommand{\ConstantTok}[1]{\textcolor[rgb]{0.56,0.35,0.01}{#1}}

\newcommand{\DecValTok}[1]{\textcolor[rgb]{0.00,0.00,0.81}{#1}}

\newcommand{\FunctionTok}[1]{\textcolor[rgb]{0.13,0.29,0.53}{\textbf{#1}}}

\newcommand{\NormalTok}[1]{#1}

\newcommand{\OtherTok}[1]{\textcolor[rgb]{0.56,0.35,0.01}{#1}}

\newcommand{\SpecialCharTok}[1]{\textcolor[rgb]{0.81,0.36,0.00}{\textbf{#1}}}

\newcommand{\StringTok}[1]{\textcolor[rgb]{0.31,0.60,0.02}{#1}}

\usepackage{graphicx}
\makeatletter
\def\maxwidth{\ifdim\Gin@nat@width>\linewidth\linewidth\else\Gin@nat@width\fi}
\def\maxheight{\ifdim\Gin@nat@height>\textheight\textheight\else\Gin@nat@height\fi}
\makeatother
\setkeys{Gin}{width=\maxwidth,height=\maxheight,keepaspectratio}
\makeatletter
\def\fps@figure{htbp}
\makeatother
\providecommand{\tightlist}{%
  \setlength{\itemsep}{0pt}\setlength{\parskip}{0pt}}
\newlength{\cslhangindent}
\newlength{\csllabelwidth}
\newlength{\cslentryspacingunit} 
\newenvironment{CSLReferences}[2] 
 {
  \setlength{\parindent}{0pt}
  \ifodd #1
  \let\oldpar\par
  \def\par{\hangindent=\cslhangindent\oldpar}
  \fi
  \setlength{\parskip}{#2\cslentryspacingunit}
 }%
 {}
\usepackage{calc}

\ifLuaTeX
  \usepackage{selnolig}  
\fi
\IfFileExists{bookmark.sty}{\usepackage{bookmark}}{\usepackage{hyperref}}
\IfFileExists{xurl.sty}{\usepackage{xurl}}{} 
\hypersetup{
  pdftitle={multilevLCA: An R Package for Single-Level and Multilevel Latent Class Analysis with Covariates},
  pdfauthor={Johan Lyrvall; Roberto Di Mari; Zsuzsa Bakk; Jennifer Oser; Jouni Kuha},
  hidelinks,
  pdfcreator={LaTeX via pandoc}}

\title{multilevLCA: An R Package for Single-Level and Multilevel Latent
Class Analysis with Covariates}
\author{Johan Lyrvall\footnote{University of Catania, Department of
  Business and Economics,
  \href{mailto:johan.lyrvall@phd.unict.it}{\nolinkurl{johan.lyrvall@phd.unict.it}},
  orcid 0000-0002-1863-8147} \and Roberto Di Mari\footnote{University of
  Catania, Department of Business and Economics,
  \href{mailto:roberto.dimari@unict.it}{\nolinkurl{roberto.dimari@unict.it}},
  orcid 0000-0001-5498-009X} \and Zsuzsa Bakk\footnote{Leiden
  University, Department of Methodology and Statistics,
  \href{mailto:z.bakk@fsw.leidenuniv.nl}{\nolinkurl{z.bakk@fsw.leidenuniv.nl}},
  orcid 0000-0001-9352-4812} \and Jennifer Oser\footnote{Ben-Gurion
  University, Department of Politics and Government,
  \href{mailto:oser@post.bgu.ac.il}{\nolinkurl{oser@post.bgu.ac.il}},
  orcid 0000-0002-1531-4606} \and Jouni Kuha\footnote{London School of
  Economics, Department of Statistics,
  \href{mailto:j.kuha@lse.ac.uk}{\nolinkurl{j.kuha@lse.ac.uk}}, orcid
  0000-0002-1156-8465}}
\date{2024-04-10}

\begin{document}
\maketitle
\begin{abstract}
This contribution presents a guide to the R package \emph{multilevLCA},
which offers a complete and innovative set of technical tools for the
latent class analysis of single-level and multilevel categorical data.
We describe the available model specifications, mainly falling within
the \emph{fixed-effect} or \emph{random-effect} approaches. Maximum
likelihood estimation of the model parameters, enhanced by a refined
initialization strategy, is implemented either simultaneously, i.e., in
one-step, or by means of the more advantageous two-step estimator. The
package features i) semi-automatic model selection when a priori
information on the number of classes is lacking, ii) predictors of class
membership, and iii) output visualization tools for any of the available
model specifications. All functionalities are illustrated by means of a
real application on citizenship norms data, which are available in the
package.
\end{abstract}

\hypertarget{introduction}{%
\section{Introduction}\label{introduction}}

Latent class (LC) analysis is used to classify units into discrete types
based on a set of observed variables which are treated as indicators of
an unknown nominal variable with some number of latent classes
(Lazarsfeld and Henry 1968; Goodman 1974; Haberman 1979). For example,
in our package illustration, we use survey responses to identify
citizenship norms among adolescents.

The LC model assumes that observations are independent of each other.
However, this assumption is often violated when the data have a
multilevel structure, namely when lower-level units are nested in
higher-level ones. In such circumstances - e.g., survey respondents
nested within countries, pupils within schools - one may reasonably
expect units belonging to the same high-level group to display more
similar scores to each other than units from different groups.

The baseline LC model specification can be extended to account for
multilevel structures in two possible ways, the \emph{fixed-effect} and
\emph{random-effect} approaches. Fixed-effect multilevel modeling
consists in single level regression analysis of the individual
clustering on covariates, including a set of dummy variables for the
higher-level groups (e.g.~countries in cross-national applications) as
covariates. This approach is particularly useful when the number of
high-level units is from small to moderate.

In the random-effect approach, the more complex dependencies are
flexibly modeled by means of a second \emph{categorical} latent variable
defined at the group level (Aitkin 1999; Laird 1978; Vermunt 2003). Such
a \emph{nonparametric} specification of the random effect model allows
us to 1) work with mild assumptions on the distribution of the random
effect, and 2) extract a clustering at higher level. The resulting
multilevel LC model can be further extended by including predictors of
class membership at both levels.

Identifying the association between class membership and external
variables (covariates) is of crucial interest in applied sciences. The
general recommendation for fitting (single-level and multilevel) LC
models with covariates is to use stepwise estimators (see, among others,
Bakk et al. 2022; Bakk, Tekle, and Vermunt 2013; Bakk and Kuha 2018; Di
Mari et al. 2023; Lanza, Tan, and Bray 2013; Vermunt 2010), which
separate the estimation of the \emph{measurement} model of the latent
classes from the estimation of the \emph{structural} part - i.e., the
portion of the model where predictors are related to class membership.
In particular, in the two-step (Di Mari et al. 2023) and two-stage
approaches (Bakk et al. 2022) for multilevel LCA, and the two-step
approach for single-level LCA (Bakk and Kuha 2018), the parameters of
the measurement model are estimated first, and then held fixed when the
full model log-likelihood is maximized only with respect to structural
parameters. General asymptotic properties of two-step estimators can be
derived from pseudo-ML theory (Gong and Samaniego 1981). In finite
samples, they have been shown to be approximately unbiased, with almost
no loss of efficiency compared to the simultaneous (one-step) maximum
likelihood estimator of all the parameters together (Di Mari et al.
2023), yet with enhanced algorithmic stability and improved speed of
convergence.

This paper describes an efficient software solution for single-level and
multilevel latent class analysis for categorical data. Both fixed-effect
and random-effect approaches are implemented for multilevel models, with
more emphasis on the latter due to its attractive properties.
\emph{multilevLCA} is the first package within the R environment
implementing simultaneous and two-step estimation, in a functional and
user-friendly way, for multilevel LC modeling of categorical data with
covariates. In addition, the package features semi-automatic model
selection, when a priori information on the number of classes is not
available. A class-profile visualization tool is also available for any
model specification.

Currently other available LCA packages in R are more limited in scope
and usability, for example \emph{poLCA} allows for single-level LCA
only, based on the simultaneous estimator; similarly, \emph{stepmixr}
focuses only on single-level LC model estimation (and model selection),
allowing for stepwise estimation yet without standard error computation.
\emph{MultiLCIRT}, although allowing for single-level LCA, is more
specialized on specifications that are commonly used in applications of
Item Response Theory (IRT) modelling. A package that is perhaps closer
in scope is \emph{glca}, which is devoted to multilevel (fixed- and
random-effect) LCA. The latter, however, allows only for a pre-specified
number of classes, covers only dichotomous items, and, perhaps most
importantly, does not include any stepwise estimator. All in all,
neither of these packages offers the combination of features that
\emph{multilevLCA} does to fit multilevel latent class models with
covariates.

The remainder of the manuscript is organized as follows. In the next
section, we present all available model specifications. We restrict the
presentation to the dichotomous item case, although we note that the
package is designed for general polytomous responses. Then, we describe
possible estimation strategies, including issues of class selection and
initialization. In the core of the manuscript, we illustrate all key
features of the package by means of a data set on adolescent's
citizenship norms, taken from the International Civic and Citizenship
Education Study 2016 (Schulz et al. 2018) of the International
Association for the Evaluation of Educational Achievement (IEA). We
conclude with some final remarks.

\hypertarget{model-specifications}{%
\section{Model specifications}\label{model-specifications}}

Let \(\small Y_{ih}\) denote the response of unit \(\small i\) on the
dichotomous item \(\small h=1,\dots,H\), and let
\(\small \boldsymbol{Y}_i=(Y_{i1},\dots,Y_{iH})^{\prime}\) denote the
full response vector for the same unit. The elements of the vector are
treated as observed indicators of the categorical latent variable
\(\small X_i\), with possible values \(\small \{1,\dots,T\}\).

The single-level latent class model for \(\small \boldsymbol{Y_i}\)
without covariates can be specified as

\begin{equation}\label{eq:eq1}
\small P(\boldsymbol{Y}_i) = \sum_{t=1}^{T} P(X_i=t) P(\boldsymbol{Y}_i \vert X_i=t).
\end{equation}

The quantity \(\small P(X_i = t)\) is the probability that unit
\(\small i\) belongs to latent class \(\small t\), and the term
\(\small P(\boldsymbol{Y}_i \vert X_i=t)\) denotes the conditional
probability of a particular response pattern \(\small \boldsymbol{Y}_i\)
given membership in class \(\small t\). With binary items, with possible
values \(\small y_{ih}=0,1\), it is natural to assume a conditional
Bernoulli distribution for the indicators, with probability mass
function
\(\small P(Y_{ih} = y_{ih} \vert X_i = t) = \phi_{h\vert t}^{y_{ih}}(1-\phi_{h\vert t})^{1-y_{ih}}\).
The latent class probabilities and the conditional response
probabilities define the latent class profiles. Responses of the
different indicators are assumed to be conditionally independent from
each other given class membership (the \emph{local independence
assumption}), leading to

\begin{equation}\label{eq:eq2}
\small P(\boldsymbol{Y}_i \vert X_i=t) = \prod_{h=1}^{H} P(Y_{ih} \vert X_i=t).
\end{equation}

In a multilevel LC model we take the lower-level units (e.g.~individual
respondents) \(\small i=1,\dots,n_j\) to be nested in higher-level units
(groups, e.g.~countries) \(\small j=1,\dots,J\). With a random-effect
specification, let \(\small W_j\) be a high-level categorical latent
variable with possible categories \(\small m=1,\dots,M\), and
probabilities \(\small P(W_j = m) = \omega_m > 0\), and let
\(\small X_{ij}\) now be a low-level categorical latent variable that is
defined conditional on the values of \(\small W_j\), with possible
values \(\small t=1,\dots,T\), and conditional probabilities
\(\small P(X_{ij} = t \vert W_j = m) = \pi_{t \vert m} > 0\). We collect
all \(\small \omega_m\) and \(\small \pi_{t \vert m}\) respectively in
the \(\small M\)-vector \(\small \boldsymbol{\omega}\), and the
\(\small M \times T\) matrix \(\small \boldsymbol{\Pi}\).

A multilevel (random effect) LC model without covariates for
\(\small \boldsymbol{Y}_{ij}\) can be specified as

\begin{equation}\label{eq:eq3}
\small P(\boldsymbol{Y}_{ij}) = \sum_{m=1}^{M} P(W_j = m) \sum_{t=1}^{T} P(X_{ij} = t \vert W_j = m) \prod_{h=1}^{H} P(Y_{ijh} \vert X_{ij}=t),
\end{equation}

where we assume that conditional response probabilities depend on
high-level class membership only through \(\small X_{ij}\). This
assumption is typically adopted in multilevel LCA to enhance model
interpretation (Vermunt 2003; Lukociene, Varriale, and Vermunt 2010).
The model specified in Equation \ref{eq:eq3} is similar to the
multilevel item response model of Gnaldi, Bacci, and Bartolucci (2016),
but with categorical latent variables at both levels.

High-level and low-level covariates can be included to predict latent
class membership. Let
\(\small \boldsymbol{Z}_{ij}=(1,\boldsymbol{Z}_{1j}^{\prime},\boldsymbol{Z}_{2ij}^{\prime})^{\prime}\)
be a vector of \(\small K\) covariates, which can be defined at the high
level (\(\small \boldsymbol{Z}_{1j}\)) and low level
(\(\small \boldsymbol{Z}_{2ij}\)). At the lower level, we consider the
following multinomial logistic model

\begin{equation}\label{eq:eq4}
\small P(X_{ij}=t \vert W_j=m,\boldsymbol{Z_{ij}}) = \frac{\exp (\boldsymbol{\gamma}_{tm}^{\prime} \boldsymbol{Z}_{ij})}{1+\sum_{s=2}^{T}\exp (\boldsymbol{\gamma}_{sm}^{\prime} \boldsymbol{Z}_{ij})},
\end{equation}

where \(\small \boldsymbol{\gamma}_{tm}\) is a vector of regression
coefficients for each \(\small t=2,\dots,T\), and
\(\small m=1,\dots,M\), and the first low-level class is taken as
reference. When only the intercept term is included, so that
\(\small \boldsymbol{Z}_{ij}=1\), then
\(\small \boldsymbol{\gamma}_{tm}\) is the log-odds
\(\small \log (\pi_{t \vert m} / \pi_{1 \vert m})\).

At higher level, the membership probabilities can be parametrized in an
analogous way, now with
\(\small \boldsymbol{Z}_{j}^{*}=(1,\boldsymbol{Z}_{1j}^{\prime})^{\prime}\)

\begin{equation}\label{eq:eq5}
\small P(W_j=m \vert \boldsymbol{Z}_{j}^{*}) = \frac{\exp (\boldsymbol{\alpha}_{m}^{\prime} \boldsymbol{Z}_{j}^{*})}{1+\sum_{l=2}^{M}\exp (\boldsymbol{\alpha}_{l}^{\prime} \boldsymbol{Z}_{j}^{*})},
\end{equation}

where \(\small \boldsymbol{\alpha}_{m}\) are regression coefficients for
\(\small m=2,\dots,M\). When only the intercept term is included, then
\(\small \boldsymbol{\alpha}_{m}\) is the log-odds
\(\small \log (\omega_m / \omega_1)\).

We further assume that the observed indicators \(\small Y_{ih}\) are
conditionally independent from the covariates given low-level class
membership. With these assumptions, the model for
\(\small P(\boldsymbol{Y}_{ij} \vert \small \boldsymbol{Z}_{ij})\) can
be written as

\begin{equation}\label{eq:eq6}
\begin{aligned}
& \small P(\boldsymbol{Y}_{ij} \vert \boldsymbol{Z}_{ij}) \\ & \small = \sum_{m=1}^{M} P(W_j = m \vert \boldsymbol{Z}_{j}^{*}) \sum_{t=1}^{T} P(X_{ij} = t \vert W_j = m,\boldsymbol{Z}_{ij}) \prod_{h=1}^{H} P(Y_{ijh} \vert X_{ij}=t).
\end{aligned}
\end{equation}

The next section describes the approaches for estimating Equation
\ref{eq:eq6} that are available in the package \emph{multilevLCA}.

\hypertarget{methodology}{%
\section{Methodology}\label{methodology}}

Let
\(\small \boldsymbol{Y}_j=(\boldsymbol{Y}_{1j},\dots,\boldsymbol{Y}_{n_jj})^{\prime}\)
denote the full set of responses for all low-level units belonging to
high-level unit \(\small j\).\footnote{ The current version of
  \emph{multilevLCA} assumes there are no missing values, and rows with
  missings on the items, if any, are removed before estimation. If
  covariates are included, rows with missing values in the predictors
  are removed only in the estimation of the structural part of the LC
  model, i.e.~(see below) step 2 in the two-step estimator, stage 2 in
  the two-stage estimator, or the single step in the one-step estimator.}

Let \(\small \boldsymbol{\Phi}\) be an \(\small H \times T\) matrix
collecting all item response probabilities \(\small \phi_{h \vert t}\),
and \(\small \boldsymbol{\Gamma}\) the \(\small (T-1)M \times K\) matrix
of (low-level) structural regression coefficients
\(\small \boldsymbol{\gamma}_{tm}\). For ease of exposition, we will
below present the methodology only for the case when
\(\small \boldsymbol{Z}_{j}^{*}=1\), so that
\(\small \boldsymbol{\alpha}_m = \log (\omega_m / \omega_1)\). However,
the case of covariates for the higher-level class is also implemented in
the package (and illustrated by the example below).

Let \(\small \boldsymbol{\theta}_1 = \text{vec} (\boldsymbol{\Phi})\) be
the measurement model parameter vector, and
\(\small \boldsymbol{\theta}_2 = (\text{vec} (\boldsymbol{\Gamma})^{\prime},\boldsymbol{\omega}^{\prime})^{\prime}\)
the structural model parameter vector, and let
\(\small \boldsymbol{\theta} = (\boldsymbol{\theta}_1^{\prime},\boldsymbol{\theta}_2^{\prime})^{\prime}\)
be the vector of all model parameters.

The default estimator in \emph{multilevLCA} is the two-step estimator
(Di Mari et al. 2023), which computes the estimates
\(\small \widetilde{\boldsymbol{\theta}}\) in the following two steps:

\begin{enumerate}
\def\labelenumi{\arabic{enumi}.}
\tightlist
\item
  Step 1 - Measurement model

  \begin{itemize}
  \tightlist
  \item
    Fit a multilevel latent class model without covariates, maximizing
    the log-likelihood \begin{equation}\label{eq:eq7}
     \small \ell (\boldsymbol{\Phi},\boldsymbol{\Pi},\boldsymbol{\omega}) = \sum_{j=1}^{J} \log P(\boldsymbol{Y}_j),
     \end{equation} by means of the EM algorithm. To initialize the EM
    algorithm,

    \begin{itemize}
    \item
      Cluster the pooled data using \(\small K\)-modes or, which is the
      default in \emph{multilevLCA}, \(\small K\)-means, with
      \(\small K=T\), on the minimum number of principal components such
      that at least 85\% of the total variance is explained (or half of
      all principal components, if this is a greater number). Use the
      resulting clustering partition to initialize an EM algorithm to
      fit a single-level \(\small T\)-class latent class model.
      Initialize \(\small \boldsymbol{\Phi}\) using the estimated
      response probabilities. Compute \(\small \widetilde{X}_{ij}\) as
      the maximum posterior class membership probability of \(\small i\)
      within \(\small j\).
    \item
      Aggregate the above posterior class membership probabilities over
      lower level individuals to obtain a \(\small J \times T\) matrix
      of relative sizes. Cluster the latter with \(\small K\)-means
      (default), or \(\small K\)-modes, with \(\small K=M\). Use the
      \(\small M\) relative class sizes to initialize
      \(\small \boldsymbol{\omega}\). Let
      \(\small \widetilde{W}_{ij}=\widetilde{W}_j\), for all
      \(\small i=1,\dots,n_j\), with \(\small \widetilde{W}_j\) being
      the \(\small K\)-means (or \(\small K\)-modes) clustering
      partition.
    \item
      Compute the \(\small T \times M\) table of relative conditional
      frequencies
      \(\small \widetilde{\boldsymbol{X}} \vert \widetilde{\boldsymbol{W}}\)
      by cross-tabulating
      \(\small \widetilde{\boldsymbol{X}} = (\widetilde{X}_{11},\dots,\widetilde{X}_{n_jJ})^{\prime}\),
      and
      \(\small \widetilde{\boldsymbol{W}} = (\widetilde{W}_{11},\dots,\widetilde{W}_{n_jJ})^{\prime}\).
      Initialize \(\small \boldsymbol{\Pi}\) using
      \(\small \widetilde{\boldsymbol{X}} \vert \widetilde{\boldsymbol{W}}\).
      The EM estimates \(\small \widetilde{\boldsymbol{\theta}}_1\)
      serve as the step 1 estimates of \(\small \boldsymbol{\theta}_1\).
    \end{itemize}
  \end{itemize}
\item
  Step 2 - Structural model

  \begin{itemize}
  \tightlist
  \item
    Fit a multilevel structural latent class model with the measurement
    parameters fixed at their step 1 estimates
    \(\small \widetilde{\boldsymbol{\theta}}_1\), maximizing the
    log-likelihood \begin{equation}\label{eq:eq8}
     \small \ell (\boldsymbol{\theta}_2 \vert \boldsymbol{\theta}_1 = \widetilde{\boldsymbol{\theta}}_1) = \sum_{j=1}^{J} \log P(\boldsymbol{Y}_j \vert \boldsymbol{Z}_{ij})
     \end{equation} with respect to \(\small \boldsymbol{\theta}_2\) by
    means of the EM algorithm. To initialize the EM algorithm,

    \begin{itemize}
    \tightlist
    \item
      Initialize \(\small \boldsymbol{\omega}\) using the step 1
      estimate \(\small \widetilde{\boldsymbol{\omega}}\).
    \item
      Initialize the intercepts \(\small \gamma_{0tm}\), for all
      \(\small m=1,\dots,M\), and \(\small t=2,\dots,T\), using the step
      1 estimate
      \(\small \log (\widetilde{\pi}_{t \vert m} / \widetilde{\pi}_{1 \vert m})\).
    \item
      Initialize the other elements of \(\small \boldsymbol{\Gamma}\) at
      zero.
    \end{itemize}
  \end{itemize}
\end{enumerate}

In addition to the two-step estimator, \emph{multilevLCA} implements the
one-step and two-stage estimators. The one-step estimator involves
estimating all the parameters simultaneously, maximizing the
log-likelihood

\begin{equation}\label{eq:eq9}
\small \ell (\boldsymbol{\theta}) = \sum_{j=1}^{J} \log P(\boldsymbol{Y}_j \vert \boldsymbol{Z}_{ij}).
\end{equation}

The two-stage approach involves computing
\(\small \widetilde{\boldsymbol{\theta}}\) in two stages. The first
stage is broken down into three steps. First (1a), fitting a
single-level latent class model without covariates, ignoring the
multilevel structure of the data. Second (1b), fitting a multilevel
latent class model without covariates, keeping the measurement model
parameters fixed at their estimated step 1a values. In case of
interaction effects between higher and lower levels, a multilevel LC
model is re-estimated (1c), keeping the high level class proportions
fixed at their step 1b values. Stage 2 of the two-stage approach is the
same as the second step of two-step estimation.

In all of these methods of estimation, the number of low-level and
high-level classes is given. Selecting these values is a distinct, yet
fundamental task. \emph{multilevLCA} implements the following model
selection strategies using the two-step approach with default estimator
settings:

\begin{itemize}
\tightlist
\item
  Sequential model selection (Lukociene, Varriale, and Vermunt 2010)

  \begin{itemize}
  \tightlist
  \item
    The low and high number of classes are selected one at a time,
    without covariates, in three steps:

    \begin{enumerate}
    \def\labelenumi{\arabic{enumi}.}
    \tightlist
    \item
      Fit a set of single-level models, selecting the optimal number of
      low-level classes.
    \item
      Fit a set of multilevel models with the number of low-level
      classes held fixed at the value of step 1, selecting the optimal
      number of high-level classes.
    \item
      Fit a set of multilevel models with the number of high-level
      classes held fixed at the value of step 2, re-selecting the
      optimal number of low-level classes.
    \end{enumerate}
  \end{itemize}
\item
  Simultaneous model selection

  \begin{itemize}
  \tightlist
  \item
    The low and high number of classes are selected simultaneously,
    without covariates, by means of multi-core estimation of all
    plausible values at both levels.
  \end{itemize}
\end{itemize}

The low-level BIC statistic is used to identify the optimal number of
low-level classes in the first and third step of the sequential
approach, and in the simultaneous approach. The high-level BIC is used
to identify the optimal number of high-level classes in the second step
of the sequential approach. Also the AIC, the low-level ICL BIC and the
high-level ICL BIC are reported in both model selection strategies.

\hypertarget{illustration-of-the-package}{%
\section{Illustration of the
package}\label{illustration-of-the-package}}

In this section we illustrate the functionalities of \emph{multilevLCA}
by estimating and visualizing a single-level latent class model without
covariates, and a multilevel latent class model with covariates at both
levels, using data on citizenship norms among adolescents in different
countries. These are the least and most complex specifications of the
available models. For simplicity of illustration, we estimate models
with a small number of classes.

We also include the fixed-effect multilevel model, which uses the same
syntax as the single-level model with covariates. Furthermore, we show
how to perform sequential and simultaneous model selection. The features
we present in this section are common to all other available
specifications.

\hypertarget{data}{%
\subsection{Data}\label{data}}

\emph{multilevLCA} includes the data set \texttt{dataIEA}, which is
taken from the International Civic and Citizenship Education Study 2016
(Schulz et al. 2018) of the IEA. As part of a comprehensive evaluation
of education systems, the IEA conducted surveys in school classes of
14--year olds to investigate civic education. The survey lists a variety
of activities for respondents to rate in terms of importance in order to
be considered a good adult citizen. Examples of questionnaire items
include obeying the law and voting in elections, protecting the
environment and defending human rights. The same data have been analyzed
in prior research (Hooghe and Oser 2015; Hooghe, Oser, and Marien 2016;
Oser and Hooghe 2013; Oser et al. 2022; Di Mari et al. 2023). All
cleaning steps to obtain the final data matrix are described {[}data
link to be made available upon publication{]}.

The answer options ``very important'' and ``quite important'' are here
coded as 1, and ``not very important'' and ``not important at all'' are
coded as 0. The covariates are individual-level socio-economic measures
and country-level measures of gross domestic product (GDP) per capita.
GDP values have been obtained from the International Monetary Fund (IMF)
website.

\begin{Shaded}
\begin{Highlighting}[]
\SpecialCharTok{\textgreater{}} \FunctionTok{str}\NormalTok{(dataIEA[,}\SpecialCharTok{{-}}\FunctionTok{c}\NormalTok{(}\DecValTok{1}\NormalTok{,}\DecValTok{3}\SpecialCharTok{:}\DecValTok{4}\NormalTok{,}\DecValTok{18}\SpecialCharTok{:}\DecValTok{19}\NormalTok{,}\DecValTok{21}\SpecialCharTok{:}\DecValTok{25}\NormalTok{,}\DecValTok{27}\SpecialCharTok{:}\DecValTok{28}\NormalTok{)])}
\end{Highlighting}
\end{Shaded}

\begin{verbatim}
## 'data.frame':    90221 obs. of  16 variables:
##  $ COUNTRY         : chr  "BGR" "BGR" "BGR" "BGR" ...
##  $ obey            : int  1 1 1 1 1 1 1 1 1 1 ...
##  $ rights          : int  1 1 1 1 1 1 1 1 1 1 ...
##  $ local           : int  1 1 1 1 1 1 1 1 1 1 ...
##  $ work            : int  1 1 1 1 1 1 1 1 1 1 ...
##  $ envir           : int  1 1 1 1 1 1 1 1 1 1 ...
##  $ vote            : int  1 1 1 1 1 1 1 1 1 1 ...
##  $ history         : int  1 1 1 1 1 0 0 1 1 1 ...
##  $ respect         : int  1 1 1 1 1 1 1 1 1 1 ...
##  $ news            : int  1 1 1 1 0 0 1 1 0 0 ...
##  $ protest         : int  1 1 0 0 1 0 1 0 1 1 ...
##  $ discuss         : int  0 0 0 0 0 0 1 1 0 1 ...
##  $ party           : int  0 0 0 0 0 0 0 0 1 0 ...
##  $ female          : int  1 0 1 0 0 1 0 0 0 0 ...
##  $ ed_mom          : int  1 2 2 2 2 2 1 2 1 2 ...
##  $ log_gdp_constant: num  9.8 9.8 9.8 9.8 9.8 ...
\end{verbatim}

\hypertarget{model-estimation-and-visualization}{%
\subsection{Model estimation and
visualization}\label{model-estimation-and-visualization}}

Latent class model estimation in \emph{multilevLCA} is done using the
function \texttt{multiLCA}. The single-level measurement model - the
simplest specification - requires three inputs: the dataset
(\texttt{data}), the names of the columns with the items (\texttt{Y}),
and the number of (low-level) classes (\texttt{iT}).

\begin{Shaded}
\begin{Highlighting}[]
\SpecialCharTok{\textgreater{}}\NormalTok{ data    }\OtherTok{=}\NormalTok{ dataIEA}
\SpecialCharTok{\textgreater{}}\NormalTok{ Y       }\OtherTok{=} \FunctionTok{colnames}\NormalTok{(dataIEA)[}\DecValTok{5}\SpecialCharTok{:}\DecValTok{16}\NormalTok{]}
\SpecialCharTok{\textgreater{}}\NormalTok{ iT      }\OtherTok{=} \DecValTok{3}
\SpecialCharTok{\textgreater{}}\NormalTok{ out     }\OtherTok{=} \FunctionTok{multiLCA}\NormalTok{(data, Y, iT)}
\end{Highlighting}
\end{Shaded}

In this \texttt{multiLCA} call, a 3-class model without covariates is
estimated. The items are the 12 columns, \texttt{obey} to
\texttt{party}.

A summary of the estimated model can be obtained as:

\begin{Shaded}
\begin{Highlighting}[]
\SpecialCharTok{\textgreater{}}\NormalTok{ out}
\end{Highlighting}
\end{Shaded}

\begin{verbatim}
## 
## CALL:
## multiLCA(data = data, Y = Y, iT = iT, verbose = FALSE)
## 
## SPECIFICATION:
##                       
##  Single-level LC model
## 
## ESTIMATION DETAILS:
## 
##  EMiter   LLfirst    LLlast
##     296 -463604.2 -442528.9
## 
## ---------------------------
## 
## CLASS PROPORTIONS:
##             
## P(C1) 0.3774
## P(C2) 0.5078
## P(C3) 0.1148
## 
## RESPONSE PROBABILITIES:
## 
##                  C1     C2     C3
## P(obey|C)    0.9822 0.9225 0.6995
## P(rights|C)  0.9868 0.8840 0.1842
## P(local|C)   0.9770 0.8446 0.2294
## P(work|C)    0.9430 0.8429 0.6251
## P(envir|C)   0.9865 0.9094 0.2941
## P(vote|C)    0.9771 0.7804 0.5287
## P(history|C) 0.9474 0.7981 0.4746
## P(respect|C) 0.9431 0.8128 0.5453
## P(news|C)    0.9678 0.7013 0.4416
## P(protest|C) 0.8873 0.5494 0.1886
## P(discuss|C) 0.8244 0.2578 0.1528
## P(party|C)   0.6083 0.1928 0.1328
## 
## ---------------------------
## 
## MODEL AND CLASSIFICATION STATISTICS:
##                     
## ClassErr      0.1448
## EntR-sqr      0.6472
## BIC      885488.2917
## AIC      885133.8536
\end{verbatim}

The results suggest that 51\%, 38\% and 11\% of the respondents belong
to classes 2, 1 and 3 respectively. Those belonging to class 1 have
consistently high conditional probabilities to score 1 (i.e., indicate
high importance) on any item. Class 2 has somewhat lower probabilities
than class 1 of placing importance on all of the items, with the largest
difference on items related to traditional political activities. Class 3
places relatively low importance on all the items.

In the terms of the LCA of the same data in Di Mari et al. (2023), we
will refer to the first and second class as ``maximal'' and ``engaged'',
respectively. The third class is a mix between the ``duty'' and
``subject'' class. We choose to label them ``subject'' because they do
not place marked emphasis on either category, and have distinctly lower
probabilities of assigning importance to all of the behaviors than the
other classes.

At the bottom of the output, we display the Akaike Information
Criterion, and the Bayesian Information Criterion (BIC), the proportion
of classification errors, and the entropy-based R\(\small ^2\) (Magidson
1981). The class separation is such that 14\% of the respondents in the
sample are expected to be miss-classified in the (modal) posterior class
assignment based on the class profiles. The error rate is decreased by
65\% when using the information on the response probabilities in
addition to the class proportions.

Next, we add covariates to this model. We use the variable for country
of residence, \texttt{COUNTRY}, effectively estimating a fixed-effect
multilevel model. The input argument for covariates on \(\small X_{ij}\)
is \texttt{Z}.

\begin{Shaded}
\begin{Highlighting}[]
\SpecialCharTok{\textgreater{}}\NormalTok{ out }\OtherTok{=} \FunctionTok{multiLCA}\NormalTok{(data, Y, iT, }\AttributeTok{Z =} \StringTok{"COUNTRY"}\NormalTok{, }\AttributeTok{extout =} \ConstantTok{TRUE}\NormalTok{)}
\end{Highlighting}
\end{Shaded}

We have set the argument \texttt{extout\ =\ TRUE} to obtain extensive
output. Subsequently, country-specific class proportions can be computed
by averaging the unit-specific class proportions over the units in a
given country. Below, this procedure is performed with Italy as an
example.

\begin{Shaded}
\begin{Highlighting}[]
\SpecialCharTok{\textgreater{}}\NormalTok{ unit\_class\_prop     }\OtherTok{=}\NormalTok{ out}\SpecialCharTok{$}\NormalTok{mU}
\SpecialCharTok{\textgreater{}}\NormalTok{ unit\_class\_prop\_ITA }\OtherTok{=}\NormalTok{ unit\_class\_prop[unit\_class\_prop[,}\StringTok{"COUNTRY.ITA"}\NormalTok{] }\SpecialCharTok{==} \DecValTok{1}\NormalTok{,]}
\SpecialCharTok{\textgreater{}}\NormalTok{ class\_prop\_ITA      }\OtherTok{=} \FunctionTok{apply}\NormalTok{(unit\_class\_prop\_ITA[,}\FunctionTok{c}\NormalTok{(}\StringTok{"C1"}\NormalTok{,}\StringTok{"C2"}\NormalTok{,}\StringTok{"C3"}\NormalTok{)],}\DecValTok{2}\NormalTok{,mean)}
\SpecialCharTok{\textgreater{}} \FunctionTok{round}\NormalTok{(class\_prop\_ITA,}\DecValTok{2}\NormalTok{)}
\end{Highlighting}
\end{Shaded}

\begin{verbatim}
##   C1   C2   C3 
## 0.63 0.32 0.05
\end{verbatim}

This shows that the proportion of the ``Maximal'' class is substantially
greater in Italy than in the overall population. On the other hand, the
``Subject'' labelled cluster is uncommon among Italian adolescents.

Next, we estimate the multilevel structural specification with
covariates on both levels. Since this is the random-effect approach, we
input \texttt{COUNTRY} as the high-level id (\texttt{id\_high}). As
low-level covariates, we use gender and maternal education. The natural
logarithm of price-constant GDP is used as a high-level predictor
(\texttt{Zh}). Finally, we choose to set the number of high-level
classes (\texttt{iM}) to 2. As always, columns are identified by their
name, as character, in the input data.

\begin{verbatim}
## 
## CALL:
## multiLCA(data = data, Y = Y, iT = iT, id_high = id_high, iM = iM, 
##     Z = Z, Zh = Zh, verbose = FALSE)
## 
## SPECIFICATION:
##                                                             
##  Multilevel LC model with lower- and higher-level covariates
## 
## ESTIMATION DETAILS:
## 
##  EMiter   LLfirst    LLlast
##      16 -426496.8 -426191.1
## 
## ---------------------------
## 
## GROUP PROPORTIONS (SAMPLE MEAN):
##             
## P(G1) 0.5909
## P(G2) 0.4091
## 
## CLASS PROPORTIONS (SAMPLE MEAN):
## 
##             G1     G2
## P(C1|G) 0.2726 0.5336
## P(C2|G) 0.5893 0.3910
## P(C3|G) 0.1381 0.0753
## 
## RESPONSE PROBABILITIES:
## 
##                  C1     C2     C3
## P(obey|C)    0.9816 0.9217 0.6991
## P(rights|C)  0.9850 0.8832 0.1816
## P(local|C)   0.9752 0.8435 0.2271
## P(work|C)    0.9437 0.8405 0.6249
## P(envir|C)   0.9871 0.9073 0.2914
## P(vote|C)    0.9733 0.7800 0.5283
## P(history|C) 0.9530 0.7909 0.4748
## P(respect|C) 0.9373 0.8149 0.5449
## P(news|C)    0.9675 0.6972 0.4414
## P(protest|C) 0.8840 0.5467 0.1864
## P(discuss|C) 0.8099 0.2607 0.1509
## P(party|C)   0.6100 0.1851 0.1331
## 
## ---------------------------
## 
## MODEL AND CLASSIFICATION STATISTICS:
##                       
## R2entrlow       0.6533
## R2entrhigh           1
## BIClow     852935.7972
## BIChigh    852533.6503
## ICLBIClow  906436.9698
## ICLBIChigh 852533.6503
## AIC        852480.1892
## 
## 
## ---------------------------
## 
## LOGISTIC MODEL FOR HIGHER-LEVEL CLASS MEMBERSHIP:
## 
## 
## MODEL FOR G2 (BASE G1)
## 
##                              Alpha   S.E.  Z-score   p-value
## alpha(Intercept|G2)         9.2283 0.1730  53.3517 0.0000***
## alpha(log_gdp_constant|G2) -0.9376 0.0170 -55.2259 0.0000***
## 
## 
##  *** p < 0.01, ** p < 0.05, * p < 0.1
## 
## ---------------------------
## 
## LOGISTIC MODEL FOR LOWER-LEVEL CLASS MEMBERSHIP:
## 
## 
## MODEL FOR C2 (BASE C1) GIVEN G1 
## 
##                          Gamma   S.E.  Z-score   p-value
## gamma(Intercept|C2,G1)  0.6735 0.0389  17.3320 0.0000***
## gamma(female|C2,G1)     0.2301 0.0196  11.7713 0.0000***
## gamma(ed_mom|C2,G1)    -0.0151 0.0370  -0.4085 0.6829   
## 
## 
##  *** p < 0.01, ** p < 0.05, * p < 0.1
## 
## MODEL FOR C3 (BASE C1) GIVEN G1 
## 
##                          Gamma   S.E.  Z-score   p-value
## gamma(Intercept|C3,G1) -0.4184 0.0287 -14.5975 0.0000***
## gamma(female|C3,G1)    -0.4470 0.0468  -9.5459 0.0000***
## gamma(ed_mom|C3,G1)    -0.0529 0.0253  -2.0918 0.0365** 
## 
## 
##  *** p < 0.01, ** p < 0.05, * p < 0.1
## 
## MODEL FOR C2 (BASE C1) GIVEN G2 
## 
##                          Gamma   S.E.  Z-score   p-value
## gamma(Intercept|C2,G2) -0.4003 0.0352 -11.3634 0.0000***
## gamma(female|C2,G2)     0.0986 0.0186   5.3018 0.0000***
## gamma(ed_mom|C2,G2)     0.0289 0.0497   0.5821 0.5605   
## 
## 
##  *** p < 0.01, ** p < 0.05, * p < 0.1
## 
## MODEL FOR C3 (BASE C1) GIVEN G2 
## 
##                          Gamma   S.E.  Z-score   p-value
## gamma(Intercept|C3,G2) -1.7046 0.0295 -57.7012 0.0000***
## gamma(female|C3,G2)    -0.6137 0.0500 -12.2784 0.0000***
## gamma(ed_mom|C3,G2)     0.0072 0.0399   0.1815 0.8560   
## 
## 
##  *** p < 0.01, ** p < 0.05, * p < 0.1
\end{verbatim}

\begin{Shaded}
\begin{Highlighting}[]
\SpecialCharTok{\textgreater{}}\NormalTok{ data    }\OtherTok{=}\NormalTok{ dataIEA}
\SpecialCharTok{\textgreater{}}\NormalTok{ Y       }\OtherTok{=} \FunctionTok{colnames}\NormalTok{(dataIEA)[}\DecValTok{5}\SpecialCharTok{:}\DecValTok{16}\NormalTok{]}
\SpecialCharTok{\textgreater{}}\NormalTok{ iT      }\OtherTok{=} \DecValTok{3}
\SpecialCharTok{\textgreater{}}\NormalTok{ id\_high }\OtherTok{=} \StringTok{"COUNTRY"}
\SpecialCharTok{\textgreater{}}\NormalTok{ iM      }\OtherTok{=} \DecValTok{2}
\SpecialCharTok{\textgreater{}}\NormalTok{ Z       }\OtherTok{=} \FunctionTok{c}\NormalTok{(}\StringTok{"female"}\NormalTok{,}\StringTok{"ed\_mom"}\NormalTok{)}
\SpecialCharTok{\textgreater{}}\NormalTok{ Zh      }\OtherTok{=} \StringTok{"log\_gdp\_constant"}
\SpecialCharTok{\textgreater{}}\NormalTok{ out     }\OtherTok{=} \FunctionTok{multiLCA}\NormalTok{(data, Y, iT, id\_high, iM, Z, Zh)}
\SpecialCharTok{\textgreater{}}\NormalTok{ out}
\end{Highlighting}
\end{Shaded}

It is estimated that 59\% of the countries belong to high-level class 1,
whereas the remaining share belongs to high-level class 2. The absolute
majority of the countries within high-level class 1 and high-level class
2 belong to low-level class 2 and low-level class 1, respectively. On
average, the countries in high-level class 2 have a lower GDP than the
countries in high-level class 1. This difference is statistically
significant at any conventional \(\small \alpha\)-level.

Within group-level class 1, girls are more likely to be ``engaged'' and
less likely to be ``subject'' relative to ``maximal'' than boys. In the
same group, adolescents with highly educated mothers are less likely to
be ``subject'' than ``maximal''. On the other hand, maternal education
seems to have no effect on the probability of being ``engaged'' relative
to the probability of being ``maximal''.

We observe the same direction of the gender effect within group-level
class 2, but the gender difference is more pronounced in distinguishing
``subject'' from ``maximal'', and less pronounced in distinguishing
``engaged'' from ``maximal''. Maternal education has no effect on
adolescent political norms in this high-level class.

The object \texttt{out} is a list with everything that is printed, with
the exception of the Z-scores and \(\small p\)-values. For example, the
(average) conditional low-level class proportions
\(\small \bar{\pi}_{t \vert m}\) are in the list object
\texttt{mPi\_avg}:

\begin{Shaded}
\begin{Highlighting}[]
\SpecialCharTok{\textgreater{}}\NormalTok{ out}\SpecialCharTok{$}\NormalTok{mPi\_avg}
\end{Highlighting}
\end{Shaded}

\begin{verbatim}
##                G1         G2
## P(C1|G) 0.2725871 0.53360821
## P(C2|G) 0.5893331 0.39104613
## P(C3|G) 0.1380798 0.07534566
\end{verbatim}

Much more model details can be obtained by setting the argument
\texttt{extout\ =\ TRUE}. The extensive output adds list objects such as
posterior class membership probabilities, and the variance-covariance
matrix of the parameter estimates of the structural model.

\begin{Shaded}
\begin{Highlighting}[]
\SpecialCharTok{\textgreater{}}\NormalTok{ out\_ext }\OtherTok{=} \FunctionTok{multiLCA}\NormalTok{(data, Y, iT, id\_high, iM, Z, Zh, }\AttributeTok{extout =} \ConstantTok{TRUE}\NormalTok{)}
\end{Highlighting}
\end{Shaded}

For example, the posterior high-level class membership probabilities for
the countries can be obtained as follows.

\begin{Shaded}
\begin{Highlighting}[]
\SpecialCharTok{\textgreater{}} \FunctionTok{head}\NormalTok{(}\FunctionTok{round}\NormalTok{(out\_ext}\SpecialCharTok{$}\NormalTok{mPW, }\DecValTok{2}\NormalTok{))}
\end{Highlighting}
\end{Shaded}

\begin{verbatim}
##     log_gdp_constant G1 G2
## BGR             9.80  1  0
## CHL            10.01  1  0
## COL             9.47  1  0
## DNK            10.70  1  0
## DOM             9.56  0  1
## EST            10.20  1  0
\end{verbatim}

The line of code below produces the matrix of posterior low-level class
membership probabilities after marginalizing over high-level classes.
The first row represents a female Bulgarian adolescent that scores 1 one
the first 10 items and 0 on the last 2 items and has a mother with a low
education. This individual has a posterior probability of 0.75 of
belonging to the ``engaged'' class.

\begin{Shaded}
\begin{Highlighting}[]
\SpecialCharTok{\textgreater{}} \FunctionTok{noquote}\NormalTok{(}\FunctionTok{head}\NormalTok{(out\_ext}\SpecialCharTok{$}\NormalTok{mPMsumX))}
\end{Highlighting}
\end{Shaded}

\begin{verbatim}
##      obey rights local work envir vote history respect news protest discuss
## [1,] 1    1      1     1    1     1    1       1       1    1       0      
## [2,] 1    1      1     1    1     1    1       1       1    1       0      
## [3,] 1    1      1     1    1     1    1       1       1    0       0      
## [4,] 1    1      1     1    1     1    1       1       1    0       0      
## [5,] 1    1      1     1    1     1    1       1       0    1       0      
## [6,] 1    1      1     1    1     1    0       1       0    0       0      
##      party COUNTRY female ed_mom C1                   C2               
## [1,] 0     BGR     1      1      0.247790720816449    0.752120500716108
## [2,] 0     BGR     0      2      0.296241549854792    0.703601052358176
## [3,] 0     BGR     1      2      0.0502391108806417   0.949192962481293
## [4,] 0     BGR     0      2      0.0623946798575403   0.93650246075248 
## [5,] 0     BGR     0      2      0.0315507790764378   0.967818610692772
## [6,] 0     BGR     1      2      0.000758483290285323 0.992008001928078
##      C3                  
## [1,] 8.8778467442711e-05 
## [2,] 0.00015739778703163 
## [3,] 0.000567926638065037
## [4,] 0.00110285938997975 
## [5,] 0.000630610230790724
## [6,] 0.00723351478163683
\end{verbatim}

\hypertarget{model-selection}{%
\subsection{Model selection}\label{model-selection}}

The default model selection approach is the sequential one. The
simultaneous approach is obtained instead by setting the argument
\texttt{sequential\ =\ FALSE}. The following function call produces
sequential model selection over 1:3 low-level classes, and 1:2
high-level classes:

\begin{Shaded}
\begin{Highlighting}[]
\SpecialCharTok{\textgreater{}}\NormalTok{ out\_seq }\OtherTok{=} \FunctionTok{multiLCA}\NormalTok{(data, Y, }\AttributeTok{iT =} \DecValTok{1}\SpecialCharTok{:}\DecValTok{3}\NormalTok{, id\_high, }\AttributeTok{iM =} \DecValTok{1}\SpecialCharTok{:}\DecValTok{2}\NormalTok{)}
\end{Highlighting}
\end{Shaded}

while the following function call produces simultaneous model selection:

\begin{Shaded}
\begin{Highlighting}[]
\SpecialCharTok{\textgreater{}}\NormalTok{ out\_sim }\OtherTok{=} \FunctionTok{multiLCA}\NormalTok{(data, Y, }\AttributeTok{iT =} \DecValTok{1}\SpecialCharTok{:}\DecValTok{3}\NormalTok{, id\_high, }\AttributeTok{iM =} \DecValTok{1}\SpecialCharTok{:}\DecValTok{2}\NormalTok{, }\AttributeTok{sequential =} \ConstantTok{FALSE}\NormalTok{)}
\end{Highlighting}
\end{Shaded}

The lists \texttt{out\_seq} and \texttt{out\_sim} contain the model
estimation results of the selected model of the sequential and the
simultaneous approaches, respectively, as if the selected specification
had been estimated directly.

\hypertarget{graphical-visualization}{%
\subsection{Graphical visualization}\label{graphical-visualization}}

The estimated response probabilities of any \texttt{multiLCA} model can
be visualized graphically with \texttt{plot}.

\begin{Shaded}
\begin{Highlighting}[]
\SpecialCharTok{\textgreater{}} \FunctionTok{plot}\NormalTok{(out)}
\end{Highlighting}
\end{Shaded}

\includegraphics[width=1\linewidth]{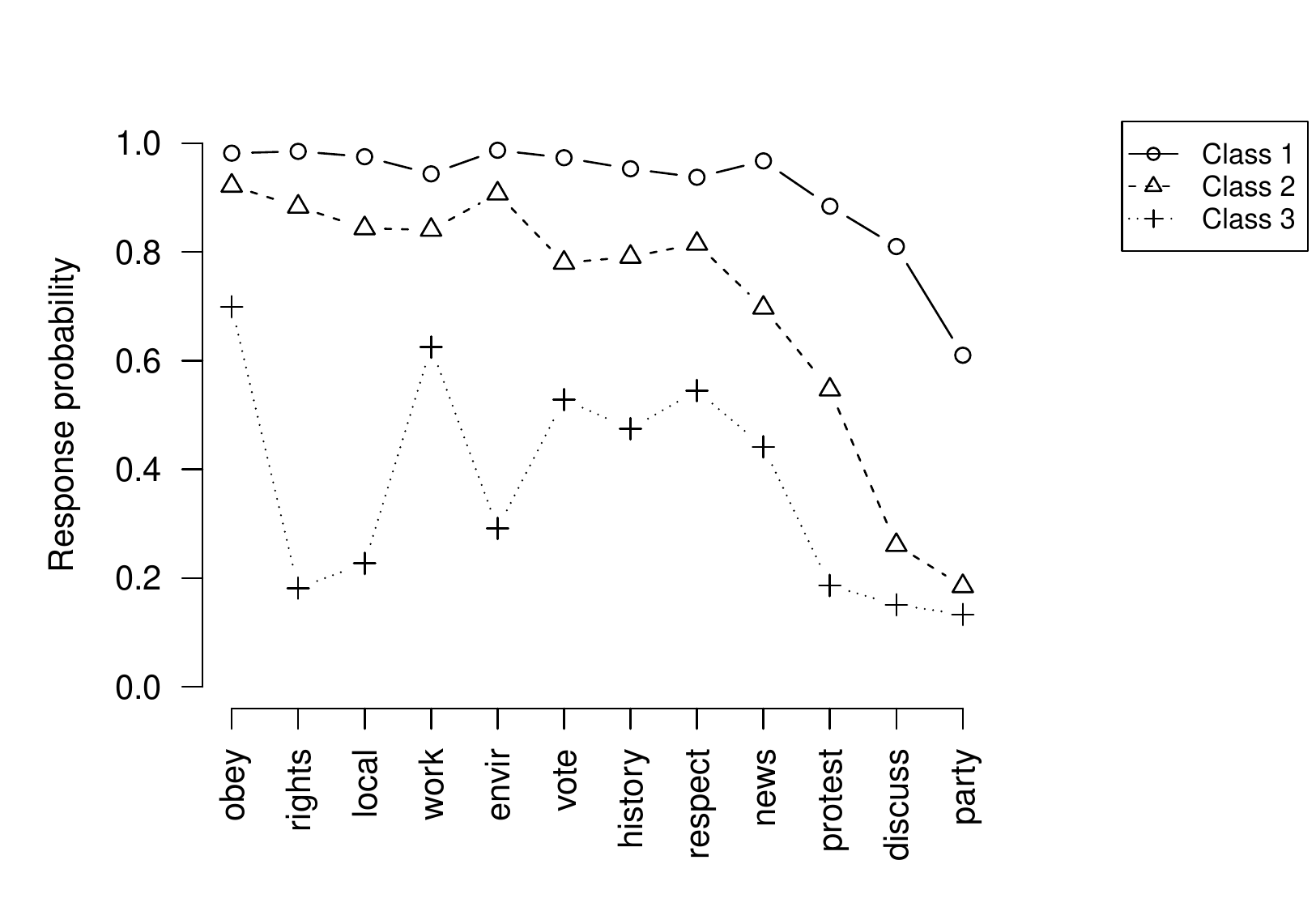}

The code below rotates the x-axis tick labels vertically, and sets the
custom labels ``Maximal'', ``Engaged'' and ``Subject''.

\begin{Shaded}
\begin{Highlighting}[]
\SpecialCharTok{\textgreater{}} \FunctionTok{plot}\NormalTok{(out\_ext, }\AttributeTok{horiz =} \ConstantTok{FALSE}\NormalTok{, }\AttributeTok{clab =} \FunctionTok{c}\NormalTok{(}\StringTok{"Maximal"}\NormalTok{, }\StringTok{"Engaged"}\NormalTok{, }\StringTok{"Subject"}\NormalTok{))}
\end{Highlighting}
\end{Shaded}

\includegraphics[width=1\linewidth]{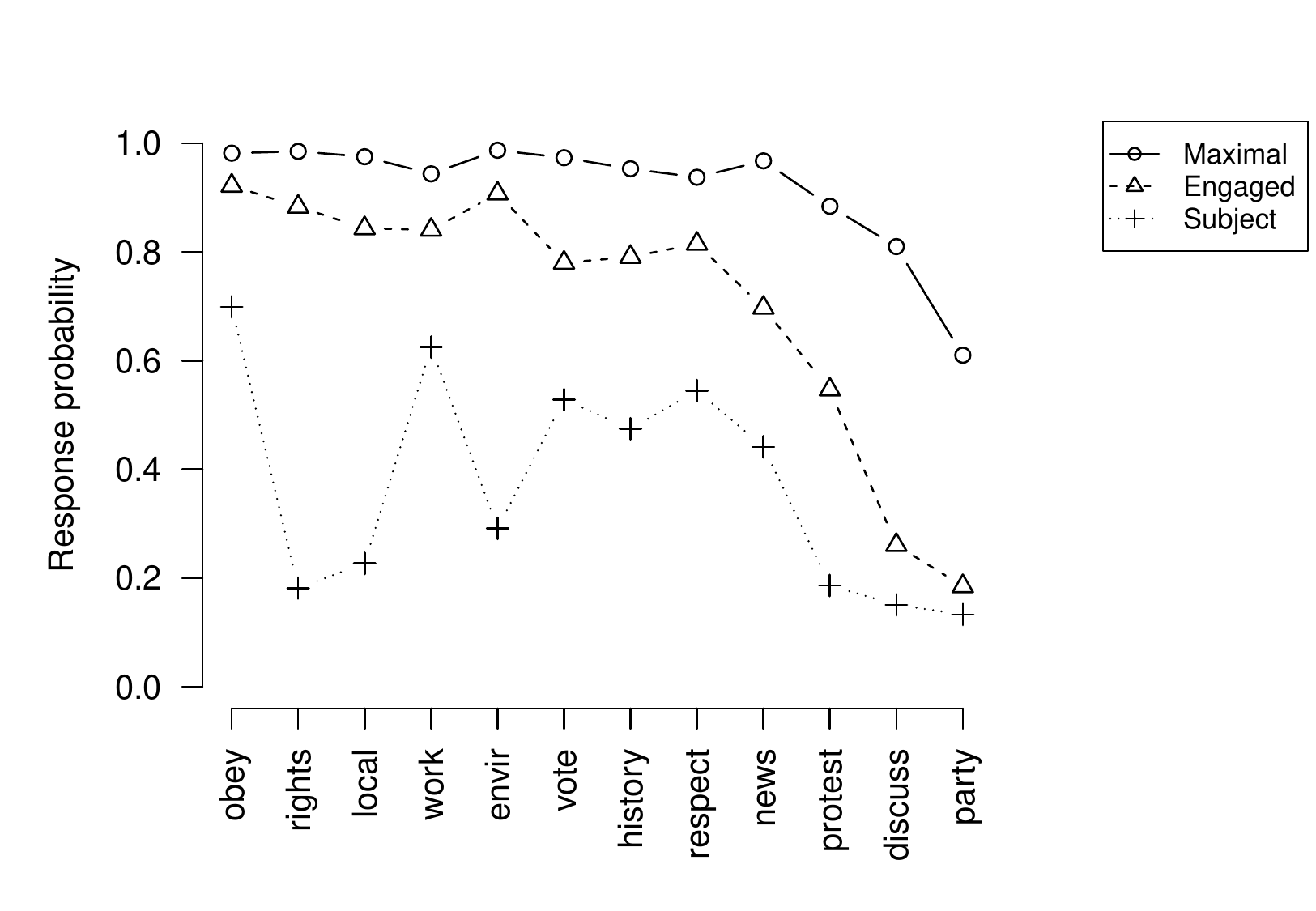}

\hypertarget{concluding-remarks}{%
\section{Concluding remarks}\label{concluding-remarks}}

We presented the latent class modelling framework that is considered by
the package \emph{multilevLCA}, and described the available methods to
efficiently compute estimates of complex LC models with a multilevel
structure and predictors of class membership at any level. We have given
a tutorial of the user-friendly syntax that executes this estimation,
visualizes the results, and implements two approaches for model
selection.

\emph{multilevLCA} fills an important gap, in that it is the first
freeware software that efficiently estimates a range of complex latent
class models, visualizes the results, and performs model selection under
the same umbrella. Thus, it offers a larger and differentiated set of
tools than available packages.

The functionalities of \emph{multilevLCA} have a wide range of
applications in fields such as the educational, political, economic,
health and behavioral disciplines. There is considerable appeal in the
unit-centric LC methodology, which allows great flexibility in the
parameterization of individual differences in a (possibly
multidimensional) phenomenon of interest.

Another approach for stepwise estimation of LC models is bias--adjusted
three--step estimation (Bolck, Croon, and Hagenaars 2004; Vermunt 2010;
see also Bakk, Tekle, and Vermunt 2013). Of the different versions of
it, the ML bias correction method of Vermunt (2010) is closer in spirit
to the two-step method on which the package focuses, as its second and
final step involves an analogous parameter fixing. For a more detailed
technical discussion on the topic, see Bakk and Kuha (2018). For this
reason, the ML method of bias-adjusted three-step estimation can be
implemented with \emph{multilevLCA}'s functions with little coding
effort.

\emph{multilevLCA} aims to disseminate the use of advanced LCA among
applied researchers from a variety of social sciences. Future versions
will consider methodologies for more detailed analysis, such as testing
for differential item functioning (Masyn 2017), and computing average
direct effects of covariates on class membership.

\hypertarget{refs}{}
\begin{CSLReferences}{1}{0}
\leavevmode\vadjust pre{\hypertarget{ref-aitkin1999}{}}%
Aitkin, M. 1999. {``A General Maximum Likelihood Analysis of Variance
Components in Generalized Linear Models.''} \emph{Biometrics} 55 (1):
117--28. \url{https://doi.org/10.1111/j.0006-341X.1999.00117.x}.

\leavevmode\vadjust pre{\hypertarget{ref-Bakketal}{}}%
Bakk, Z., R. Di Mari, J. Oser, and J. Kuha. 2022. {``Two-Stage
Multilevel Latent Class Analysis with Covariates in the Presence of
Direct Effects.''} \emph{Structural Equation Modeling: A
Multidisciplinary Journal} 29 (2): 267--77.
\url{https://doi.org/10.1080/10705511.2021.1980882}.

\leavevmode\vadjust pre{\hypertarget{ref-BakkKuha}{}}%
Bakk, Z., and J. Kuha. 2018. {``Two-Step Estimation of Models Between
Latent Classes and External Variables.''} \emph{Psychometrika} 83:
871--92. \url{https://doi.org/10.1007/s11336-017-9592-7}.

\leavevmode\vadjust pre{\hypertarget{ref-bakk2013}{}}%
Bakk, Z., F. B. Tekle, and J. K. Vermunt. 2013. {``Estimating the
Association Between Latent Class Membership and External Variables Using
Bias-Adjusted Three-Step Approaches.''} \emph{Sociological Methodology}
43 (1): 272--311. \url{https://doi.org/10.1177/0081175012470644}.

\leavevmode\vadjust pre{\hypertarget{ref-BCH}{}}%
Bolck, A., M. Croon, and J. Hagenaars. 2004. {``Estimating Latent
Structure Models with Categorical Variables: One-Step Versus Three-Step
Estimators.''} \emph{Political Analysis} 12 (1): 3--27.
\url{https://doi.org/10.1093/pan/mph001}.

\leavevmode\vadjust pre{\hypertarget{ref-DiMarietal}{}}%
Di Mari, R., Z. Bakk, J. Oser, and J. Kuha. 2023. {``A Two-Step
Estimator for Multilevel Latent Class Analysis with Covariates.''}
\emph{arXiv Preprint arXiv:2303.06091}.
\url{https://doi.org/10.48550/arXiv.2303.06091}.

\leavevmode\vadjust pre{\hypertarget{ref-Gnaldietal}{}}%
Gnaldi, M., S. Bacci, and F. Bartolucci. 2016. {``A Multilevel Finite
Mixture Item Response Model to Cluster Examinees and Schools.''}
\emph{Advances in Data Analysis and Classification} 10: 53--70.
\url{https://doi.org/10.1007/s11634-014-0196-0}.

\leavevmode\vadjust pre{\hypertarget{ref-gong1981}{}}%
Gong, G., and F. J. Samaniego. 1981. {``Pseudo Maximum Likelihood
Estimation: Theory and Applications.''} \emph{The Annals of Statistics},
861--69.

\leavevmode\vadjust pre{\hypertarget{ref-Goodman}{}}%
Goodman, L. A. 1974. {``The Analysis of Systems of Qualitative Variables
When Some of the Variables Are Unobservable. Part i: A Modified Latent
Structure Approach.''} \emph{American Journal of Sociology} 79:
1179--259. \url{https://doi.org/10.1086/225676}.

\leavevmode\vadjust pre{\hypertarget{ref-Haberman}{}}%
Haberman, S. 1979. \emph{Analysis of Qualitative Data: New
Developments}. Academic Press.

\leavevmode\vadjust pre{\hypertarget{ref-HoogheOser}{}}%
Hooghe, M., and J. Oser. 2015. {``The Rise of Engaged Citizenship: The
Evolution of Citizenship Norms Among Adolescents in 21 Countries Between
1999 and 2009.''} \emph{International Journal of Comparative Sociology}
56 (1): 29--52. \url{https://doi.org/10.1177/0020715215578488}.

\leavevmode\vadjust pre{\hypertarget{ref-Hoogheetal}{}}%
Hooghe, M., J. Oser, and S. Marien. 2016. {``A Comparative Analysis of
'Good Citizenship': A Latent Class Analysis of Adolescents' Citizenship
Norms in 38 Countries.''} \emph{International Political Science Review}
37 (1): 115--29. \url{https://doi.org/10.1177/0192512114541562}.

\leavevmode\vadjust pre{\hypertarget{ref-laird1978}{}}%
Laird, N. 1978. {``Nonparametric Maximum Likelihood Estimation of a
Mixing Distribution.''} \emph{Journal of the American Statistical
Association} 73 (364): 805--11.
\url{https://doi.org/10.1080/01621459.1978.10480103}.

\leavevmode\vadjust pre{\hypertarget{ref-Lanza13}{}}%
Lanza, T. S., X. Tan, and C. B. Bray. 2013. {``Latent Class Analysis
with Distal Outcomes: A Flexible Model-Based Approach.''}
\emph{Structural Equation Modeling: A Multidisciplinary Journal} 20:1:
1--26. \url{https://doi.org/10.1080/10705511.2013.742377}.

\leavevmode\vadjust pre{\hypertarget{ref-LazarsfeldHenry}{}}%
Lazarsfeld, P. F., and N. W. Henry. 1968. \emph{Latent Structure
Analysis}. Houghton Mifflin.

\leavevmode\vadjust pre{\hypertarget{ref-Lukociene}{}}%
Lukociene, O., R. Varriale, and J. K. Vermunt. 2010. {``The Simultaneous
Decision(s) about the Number of Lower- and Higher-Level Classes in
Multilevel Latent Class Analysis.''} \emph{Sociological Methodology} 40
(1): 247--83. \url{https://doi.org/10.1111/j.1467-9531.2010.01231.x}.

\leavevmode\vadjust pre{\hypertarget{ref-magidson81}{}}%
Magidson, J. 1981. {``Qualitative Variance, Entropy, and Correlation
Ratios for Nominal Dependent Variables.''} \emph{Social Science
Research} 10: 177--94.
\url{https://doi.org/10.1016/0049-089X(81)90003-X}.

\leavevmode\vadjust pre{\hypertarget{ref-Masyn}{}}%
Masyn, K. E. 2017. {``Measurement Invariance and Differential Item
Functioning in Latent Class Analysis with Stepwise Multiple Indicator
Multiple Cause Modeling.''} \emph{Structural Equation Modeling: A
Multidisciplinary Journal} 24 (2): 180--97.
\url{https://doi.org/10.1080/10705511.2016.1254049}.

\leavevmode\vadjust pre{\hypertarget{ref-OserHooghe}{}}%
Oser, J., and M. Hooghe. 2013. {``The Evolution of Citizenship Norms
Among {S}candinavian Adolescents, 1999-2009.''} \emph{Scandinavian
Political Studies} 36 (4): 320--46.
\url{https://doi.org/10.1111/1467-9477.12009}.

\leavevmode\vadjust pre{\hypertarget{ref-Oseretal}{}}%
Oser, J., M. Hooghe, Z. Bakk, and R. Di Mari. 2022. {``Changing
Citizenship Norms Among Adolescents, 1999-2009-2016: A Two-Step Latent
Class Approach with Measurement Equivalence Testing.''} \emph{Quality \&
Quantity}, 1--19. \url{https://doi.org/10.1007/s11135-022-01585-5}.

\leavevmode\vadjust pre{\hypertarget{ref-Schulz}{}}%
Schulz, W., J. Ainley, J. Fraillon, B. Losito, G. Agrusti, and T.
Friedman. 2018. \emph{Becoming Citizens in a Changing World: IEA
International Civic and Citizenship Education Study 2016 International
Report}. Springer Nature.

\leavevmode\vadjust pre{\hypertarget{ref-Vermunt}{}}%
Vermunt, J. K. 2003. {``Multilevel Latent Class Models.''}
\emph{Sociological Methodology} 33 (1): 213--39.
\url{https://doi.org/10.1111/j.0081-1750.2003.t01-1-00131.x}.

\leavevmode\vadjust pre{\hypertarget{ref-vermunt2010}{}}%
---------. 2010. {``Latent Class Modeling with Covariates: Two Improved
Three-Step Approaches.''} \emph{Political Analysis} 18 (4): 450--69.
\url{https://doi.org/10.1093/pan/mpq025}.

\end{CSLReferences}

\end{document}